\documentstyle[pre,aps,multicol,epsf]{revtex}
\renewcommand{\narrowtext}{\begin{multicols}{2} \global\columnwidth20.5pc}
\renewcommand{\widetext}{\end{multicols} \global\columnwidth42.5pc}
\newcommand{\Rrule}{\vspace{-0.1in}\hfill\vrule depth1em height0pt \vrule
  width3.5in height.2pt depth.2pt\vspace*{-0.125in}}
\def\beq{\begin{equation}}
\def\eeq{\end{equation}}
\def\bea{\begin{eqnarray}}
\def\eea{\end{eqnarray}}
\def\pa{\partial}
\def\e{{\rm e}}
\def\tr{{\rm tr}}
\def\K{\hat{K}}
\def\x{\hat{x}}
\def\q{\tilde{q}}
\def\p{\tilde{p}}
\def\R{\tilde{R}}
\begin{document}
\draft
\title{ 
Level spacing distribution of
critical random matrix ensembles
}
\author{Shinsuke M. Nishigaki${}^*$}
\address{
Institute for Theoretical Physics, University of California,
Santa Barbara, California 93106
}
\date{Revised 16 October 1998}
\maketitle
\begin{abstract} 
We consider unitary invariant random matrix ensembles
which obey spectral statistics different from the Wigner-Dyson,
including unitary ensembles with slowly ($\sim \log^2 x$) growing potentials
and the finite-temperature fermi gas model.
If the deformation parameters in these matrix ensembles are small,
the asymptotically translational-invariant region
in the spectral bulk is universally governed by a one-parameter 
generalization of the sine kernel.
We provide an analytic expression for the distribution of 
the eigenvalue spacings of this universal asymptotic kernel,
which is a hybrid of the Wigner-Dyson and the Poisson distributions,
by determining the Fredholm determinant of the universal kernel
in terms of a Painlev\'{e} VI transcendental function.
\end{abstract}
\narrowtext

A main goal of quantum chaos study is to describe quantitatively
statistical behaviors of spectra of classically non-integrable systems,
such as complex nuclei \cite{Por},
billiards \cite{BGS}, QCD \cite{BB}, and disordered systems \cite{Efe}.
A characteristic observable in such studies,
used analytically or numerically to measure the deviation from integrability,
is the probability $E(s)$ 
of having no energy levels
in an interval of width $s$, or
the distribution of spacings between
adjacent levels $P(s)=E''(s)$.
These observables capture the behavior of local correlations of 
large number of energy levels,
as the former consists of an infinite sum of integrals of 
regulated${}^\dagger$ spectral correlators,
\bea
&&E(s)\!=\! \sum_{p=0}^\infty 
\frac{(-1)^p}{p!} 
\int_{-s/2}^{s/2} \!\! dx_1\cdots dx_p
\left\langle \rho(x_1)\cdots\rho(x_p)\right\rangle_{\rm reg}.\!\!\!
\label{Es}
\eea
A technical virtue of 
invariant random matrix models of quantum chaotic systems \cite{Meh,GMW}
is that 
any $p$-point spectral correlation function of the former
can be composed from the connected two-point function as
\beq
\left\langle \rho(x_1)\cdots\rho(x_p)\right\rangle_{\rm reg}
=\det\limits_{1\leq i,j\leq p} K(x_i,x_j).
\label{detexp}
\eeq
This can most easily be proved by the orthogonal polynomial
method \cite{Meh}.
It allows a compact expression for the level-free probability $E(s)$
as a Fredholm determinant,
\beq
E(s)=\det(1-\K)
\eeq
over the interval $[-{s}/{2},{s}/{2}]$.

Jimbo et.\ al.\  \cite{JMMS} have made a remarkable observation
that the logarithmic derivative of the Fredholm determinant,
$R(s)= -(\log E(s))'$,
of the sine kernel
\beq
K(x,y)=\frac{\sin \pi (x-y)}{\pi(x-y)},
\label{Ksin}
\eeq
describing the bulk correlation of the Gaussian unitary ensemble,
satisfies the $\sigma$-form of a Painlev\'e V equation,
\bea
&&\left(R'(s)\!+\!\frac{s}{2}R''(s)\!\right)^2
\!\!\!+\!
(\pi s R'(s))^2
\!=\! R'(s)(R(s) \!+\! s R'(s))^2 \!\!.\!\!\!
\label{PV}
\eea
Their method is subsequently generalized by Tracy and Widom
\cite{TW4} to kernels of the form
\beq
K(x,y)=\frac{\phi(x)\psi(y)-\psi(x)\phi(y)}{x-y}
\eeq
whose component functions satisfy a set of
first order linear differential equations with meromorphic coefficients.
This class of kernels includes Airy \cite{TW2}, Bessel \cite{TW3},
and their multicritical generalizations \cite{KF,ADMN}, 
describing correlations at the edges of the spectral bands.

However, such invariant random matrix ensembles 
are sometimes crude idealization
of physical systems in concern, based only on the symmetries of the systems.
It is a priori unclear that random matrix ensembles can still provide
quantitative description of 
realistic physical systems where localization of the states can occur
\cite{SSSLS}.
If there exists such a random matrix ensemble, 
it must be a non-trivial deformation of
the classical invariant random matrix ensembles,
so as to violate the wide universality that the Gaussian
ensembles possess \cite{BZ}.
One such example is a random banded matrix ensemble \cite{CMI},
modeling quasi 1D materials.
Another example is a random hamiltonian
consisting of a Gaussian random matrix $H$
and a diagonal real random matrix $V$,
$H_{(\alpha)}=H+\alpha V$.
In such cases spectral correlation functions generally do not allow 
the determinant form (\ref{detexp}),
and $E(s)$ is usually evaluated only perturbatively in $s$,
by computing each $p$-point correlation function \cite{Guh}. 
However this is not sufficient to determine $P(s)=E''(s)$ for large $s$,
since it typically takes the form $P(s)\sim s^a \exp(-{\rm cst.} s^b)$
(generalizations of the Wigner surmise
by Brody \cite{Bro} and by Berry and Robnik \cite{BR}),
whose exponential damping is invisible in the small-$s$ expansion.
Alternatively, 
by treating evolution in $\alpha$ as a diffusion process
the joint probability distribution can be derived \cite{Haa},
but the level spacing distribution is yet to be obtained.
The aim of this Letter is to derive $P(s)$ from 
random matrix ensembles which describes a 
deformation of the Wigner-Dyson statistics
while preserving unitary invariance at the level of partition function.

Muttalib et.\ al.\ \cite{MCIN} have introduced
the $q$-analogue of the GUE
with a potential $(0<q<1)$
\beq
V(\lambda)=\sum_{n=1}^\infty \log(1+2q^n \cosh (2\,{\rm arcsinh}\,
\lambda)+q^{2n}).
\eeq
They have shown that, after unfolding the spectrum
\beq
\lambda \mapsto x=\int^\lambda \rho(\lambda)d\lambda,
\eeq
this ensemble is described by a kernel
\bea
K(x,y)&=&\frac{f(x+y)}{\sqrt{f(2x) f(2y)}}
\frac{\theta_1(\pi (x-y),\e^{-{\pi^2 \over a}})}{\sinh a(x-y)}, 
\label{KMCIN}\\
f(x)&\equiv& \frac{\theta_4(\pi x,\e^{-{\pi^2 \over a}})}{\cosh a x},
\ \ 
a\equiv\frac{1}{2}\log \frac{1}{q}.
\nonumber
\eea
For $\e^{-{\pi^2}/{a}} \ll 1$,
there exists an asymptotically translational invariant region
where (\ref{KMCIN}) is approximated by
\beq
K(x,y)=\frac{a \sin \pi (x-y)}{\pi \sinh a (x-y)}.
\label{KMNS}
\eeq
It is clear from this form of the kernel that 
a set of eigenvalues with $|x_i-x_j| \ll 1/a$ obeys the Wigner statistics, and
that with $|x_i-x_j| \gg 1/a$ obeys the 
Poisson statistics, i.e. is uncorrelated.
Canali and Kravtsov \cite{CK} argued that
the $U(N)$ symmetry is spontaneous broken in this ensemble,
which induces a preferred basis in the matrices and deforms
the statistics.
The universality of this asymptotic kernel (\ref{KMNS}) observed for
the $q$-Laguerre unitary ensemble \cite{BCM}
and subsequently proved for ensembles with potentials
which grow very slowly as 
\beq
V(\lambda)\sim \frac{1}{2a} (\log |\lambda|)^2
\ \ (\lambda \rightarrow \infty).
\label{log2}
\eeq
\noindent
$\!$\cite{CK}.
This universality can be considered as an extension of Br\'{e}zin and Zee's
universality \cite{BZ}
of the sine kernel for polynomially increasing potentials,
who proved it by deriving the asymptotic form of the wave functions
\bea
&&\psi_N(\lambda) \sim \cos\left(
\pi \int^\lambda \rho(\lambda)d\lambda + \frac{N\pi}{2} \right) , \\
&&K(\lambda,\lambda')\sim
\frac{\sin(\pi (
\int^\lambda \rho - \int^{\lambda'} \rho))}{\lambda-\lambda'}.
\label{K}
\eea
For polynomially increasing potentials, the spectral density is bounded 
and is locally approximated by a constant. Therefore the unfolding
is just a linear transformation, leading 
universally to the sine kernel (\ref{Ksin}).
However, for the potential (\ref{log2}), the spectral density behaves as
$\rho(\lambda) \sim 1/(2a\lambda)$ implying an unusual unfolding map 
$\lambda \mapsto x=1/(2a)\log \lambda$, while the formula (\ref{K})
stays valid \cite{CL}.
Then the kernel (\ref{K}) universally reduces to (\ref{KMNS})
after this unfolding.

Chen and Muttalib \cite{CM} have interpreted
a particular unitary ensemble with 
$V(x) \sim (\log x)^2$ as a fermionic system
at finite temperature.
This link is made more concrete
by Moshe, Shapiro, and Neuberger 
\cite{MNS} who have introduced 
a random matrix ensemble
\beq
Z=\int d^{N^2}H\,\e^{-\tr H^2} \int_{U(N)} dU\, \e^{-b\,\tr[U,H][U,H]^\dagger},
\label{ZMNS}
\eeq
as yet another unitary invariant deformed ensemble.
For a given unitary matrix $U$, the interaction
$b\,\tr[U,H][U,H]^\dagger$ tries to align random hermitian
matrices $H$ so that $[U,H]=0$.
The integration over $U$ then amounts to 
recovering the $U(N)$ invariance of the model which the GUE has enjoyed.
The basis preference is still realized through the spontaneous
breaking of the $U(N)$ invariance.
After integrating over $U$,
the above model is identical to
a system of 1D free nonrelativistic fermions
in a harmonic potential with curvature $m=\sqrt{1+4b}$
and at finite temperature
$\beta= {\rm arccosh}(1+{1}/{2b})$,
extending the well known fact that the GUE is equivalent
to free harmonic fermions at zero temperature.
Note that $p$-point correlation functions of the local spectral densities
$
\rho(\lambda)=
\sum_{n=0}^\infty{|\psi_n(\lambda)|^2}/({\e^{\beta(\epsilon_n-\mu)}+1})
$
are thus still expressed as determinants of the kernel
\beq
K(\lambda,\lambda')=
\sum_{n=0}^\infty
\frac{\psi_n(\lambda)\psi^*_n(\lambda')}{\e^{\beta(\epsilon_n-\mu)}+1}.
\label{Kn}
\eeq
Accordingly its level-free probability can be expressed in terms
of the Fredholm determinant of (\ref{Kn}).
Using the asymptotics of the
one-particle wave function $\psi_n(x)$
given by the Hermite polynomials,
these authors have also obtained the local form of the kernel
(\ref{KMNS}) with $a=\pi^2/(2N\beta)$.
There the limit $N\rightarrow \infty$ is taken while 
keeping the microscopic unfolded variable $x$ fixed.
The formula (\ref{KMNS}) is valid 
as long as $N\beta$ is not too high 
to invalidate the grand canonical picture.
The models
(\ref{log2}) and (\ref{ZMNS}) 
are subsequently unified as an ensemble with 
multifractal eigenvectors \cite{KM}, and 
the parameter $a$ is identified
as a measure of the multifractality.

Surprisingly, this universality within random matrix theories,
if extended to orthogonal ensembles \cite{W},
is observed to encompass the 3D Anderson model,
i.e. a particle hopping on the lattice with random cite energies.
Canali \cite{C} has compared his Monte Carlo results of
$P(s)$ for the orthogonal ensembles with $(1/2a)\log^2 x$ potentials 
with that for the Anderson Hamiltonian at the
metal/insulator transition measured precisely in
ref.\cite{ZK} by exact diagonalization, and
observed an excellent agreement by tuning the coefficient
to $a \approx 2.5$.
There $a$ is interpreted as the inverse 
dimensionless conductance at the transition point.
Motivated by this success,
we derive an analytic form of its 
level spacing distribution of 
(\ref{KMNS})
in this short Letter.
Although the $a\rightarrow \infty$ limit of the model does not obey 
the Poissonian statistics \cite{BBP} as is naively expected,
the error involved in the asymptotic kernel (\ref{KMNS}) is exponentially
small (of order $O(\e^{-{\pi^2}/{a}})$) in the above parameter range.
We complete earlier attempts which computed $P(s)$ numerically
\cite{MCIN} or asymptotically \cite{MK}.

We notice that the kernel (\ref{KMNS})
is equivalent to that of 
Dyson's circular unitary ensemble
at finite-$N$ \cite{Meh}:
\beq
K(x,y)=\frac{\sin \frac{N}2 (x-y)}{N \sin \frac12 (x-y)}
\label{KCUE}
\eeq
by the following analytic continuation:
\beq
N\rightarrow \frac{\pi i}{a},\ \ 
x\rightarrow \frac{2a}{i}x
\label{sub}
\eeq
Tracy and Widom \cite{TW4} have also proved that the diagonal resolvent kernel
of (\ref{KCUE}) is determined by
a second-order differential
equation which is reduced to a Painlev\'e VI equation
\cite{DIZ}.
Below we will reproduce their method.

The kernel (\ref{KMNS}) is written as
\bea
&&K(x,y)=\frac{\phi(x)\psi(y)-\psi(x)\phi(y)}{\e^{2ax}-\e^{2ay}},
\label{Kpp}\\
&&\phi(x)=\sqrt{\frac{2a}{\pi}}\,\e^{ax}\sin \pi x,\ \ 
\psi(x)=\sqrt{\frac{2a}{\pi}}\,\e^{ax}\cos \pi x.\nonumber
\eea
These component functions satisfy 
\beq
\phi'=a \phi + \pi \psi,\ \ 
\psi'=-\pi \phi + a \psi.
\eeq 
We use the bra-ket notation $\phi(x)=\langle x | \phi \rangle$
and so on \cite{BH}. 
Due to our choice of the component functions
to be real valued (unlike \cite{TW4}, sect.\ V.D), we have 
$\langle x | \hat{O} | \phi \rangle = \langle \phi | \hat{O} | x \rangle$
and similarly for $\psi$
with any self-adjoint operator $\hat{O}$ and real $x$.
Then (\ref{Kpp}) is equivalent to
\beq
[\e^{2a\x}, \K]=|\phi\rangle \langle \psi |-|\psi\rangle \langle \phi|,
\label{eK}
\eeq
where $\x$ and $\K$ are 
the multiplication operator
of the independent variable and the integral operator with the kernel
$K(x, y)\theta(y-t_1) \theta(t_2-y) $, respectively. 
Below we will not explicitly write the dependence 
on the end points of the underlying interval $[t_1,t_2]$.
The resolvent kernel is defined as
\beq
R(x,y)= \langle x | \frac{\K}{1-\K} | y \rangle.
\eeq
It follows from (\ref{eK}) that
\beq
[\e^{2a\x}, \frac{\K}{1-\K}]=
\frac{1}{1-\K}
( |\phi\rangle \langle \psi| - |\psi\rangle \langle \phi| )
\frac{1}{1-\K} ,
\label{eR}
\eeq
that is, 
\bea
&&(\e^{2ax}-\e^{2ay}) R(x,y)= Q(x)P(y)-P(x)Q(y),
\label{Rxy}\\
&&Q(x)\equiv\langle x | (1-\K)^{-1}| \phi \rangle,\ \ 
P(x)\equiv\langle x | (1-\K)^{-1}| \psi \rangle.
\nonumber
\eea
At a coincident point $x=y$ we have
\beq
2a\,\e^{2ax} R(x,x)= Q'(x)P(x)-P'(x)Q(x).
\label{Rxx}
\eeq
Now, by using the identity
\beq
\frac{\pa \K}{\pa {t_i}}=(-1)^{i} \K |t_i \rangle \langle t_i |,
\ \ (i=1,2)
\eeq
we obtain
\begin{mathletters}
\label{dPQdti}
\bea
&&\frac{\pa Q(x)}{\pa {t_i}}=(-1)^{i} R(x, t_i) Q(t_i),
\label{dQdti}\\
&&\frac{\pa P(x)}{\pa {t_i}}=(-1)^{i} R(x, t_i) P(t_i).
\label{dPdti}
\eea
\end{mathletters}
On the other hand, by using the identity 
($D$ is the derivation operator)
\beq
[D,\K]=\K( |t_1 \rangle \langle t_1 |- |t_2 \rangle \langle t_2 |) ,
\eeq
which follows from the translational invariance of the kernel
$(\pa_x +\pa_y) K(x-y)=0$,
we also have 
\begin{mathletters}
\label{dPQdx}
\bea
&&\frac{\pa Q(x)}{\pa x}
=\langle x| D (1-\K)^{-1} |\phi \rangle \nonumber\\
&&=\langle x| (1-\K)^{-1} |\phi' \rangle 
+\langle x| (1-\K)^{-1}[D,\K](I-\K)^{-1} |\phi \rangle
\nonumber\\
&&=a Q(x) + \pi P(x) + R(x,t_1)Q(t_1)- R(x,t_2)Q(t_2),
\label{dQdx} \!\!\\
&&\frac{\pa P(x)}{\pa x}= 
\nonumber\\
&&-\pi Q(x) + a P(x) + R(x,t_1)P(t_1)- R(x,t_2)P(t_2) .\!\!
\label{dPdx}
\eea
\end{mathletters}
Now we set $t_1=-t$, $t_2=t$, $x, y= -t$ or $t$, and introduce notations
$\q=Q(-t), q=Q(t), \p=P(-t), p=P(t)$, and
$\R=R(-t,t)=R(t,-t), R=R(t,t)=R(-t,-t)$. 
The last two equalities follow from
the evenness of the kernel.
Then eqs.(\ref{Rxy}) and (\ref{Rxx}) reads, after using (\ref{dPQdx}),
\begin{mathletters}
\label{pqpq}
\bea
&&  \p q-\q p= 2\R \sinh 2at ,\\
&&  \p^2+\q^2=\frac2\pi ( \R^2 \sinh 2at + R\, a\, \e^{-2at} ) , \\
&&   p^2+ q^2=\frac2\pi ( \R^2 \sinh 2at + R\, a\, \e^{ 2at} ) .
\eea
\end{mathletters}
The total $t$-derivatives of eqs.(\ref{pqpq}) lead to
($\cdot = d/dt$)
\bea
&&  \p p+ \q q= \frac1\pi (\R \sinh 2at)^{\mbox{$\cdot$}},
\label{ppqq}\\
&& 
\dot{R}=2\R^2, \ \ \ddot{R}=4\R \dot{\R}.
\label{RR}
\eea
The left hand sides of eqs.(\ref{pqpq}) and (\ref{ppqq})
satisfy an additional constraint
\beq
(\p p+\q q)^2+ (\p q-\q p)^2=
(\p^2 + \q^2) (p^2+q^2) .
\label{const}
\eeq
By eliminating $\p, p, \q, q$, $\tilde{R}$ and $\dot{\R}$ from 
eqs.(\ref{pqpq})-(\ref{const}), 
we finally obtain for 
$R(s)$ ($s \equiv 2t$, $'=d/ds$)
\bea
&&\left( a \cosh as\,R'(s)  \!+\! \frac{\sinh as}{2}R''(s) \right)^2
\!\! + \! (\pi \sinh as\,R'(s))^2 =
\nonumber\\
&&R'(s)\left( (a R(s))^2 \!+\!a\sinh 2as R(s)R'(s)
\!+\! (\sinh as\,R'(s))^2 \right)\!.
\nonumber\\
&&
\eea
\noindent
This is our main result. 
It is equivalent to eq.(5.70) of ref.\cite{TW4} after the
analytic continuation
(\ref{sub}) accompanied by a redefinition
$R(s)\rightarrow \frac{i}{2a} R(s)$.${}^\ddagger$
It clearly reduces to the Painlev\'{e} V equation (\ref{PV}) for the GUE
as $a\rightarrow 0$.
In the Wigner-like region $as\ll 1$,
we can expand 
hyperbolic functions into the Taylor series.
Then (\ref{Es}) and (\ref{detexp}) yield
\bea
E(s)&=& 1-s+O(s^4), \nonumber \\
R(s)&=& -(\log E(s))'=1+s+O(s^2)
\label{bc},
\eea
By imposing this boundary condition,
we obtain a perturbative solution to eq.(33),
\widetext
\bea
R(s)&\!=\!&
1 + s + s^2
+ \left(1-\frac{\pi^2\!+\!a^2}{9}\right) s^3  
+ \left(1-\frac{5(\pi^2\!+\!a^2)}{36}\right) s^4  
+ \left(1-\frac{(\pi^2+a^2)(75 -4\pi^2-6a^2)}{450}\right) s^5
\!+\! \cdots,\!\!\!\\
P(s)&\!=\!&
\left( R(s)^2-R'(s) \right) \exp \left({ -\int_0^s ds\,R(s) }\right)
\nonumber\\
&\!=\!&\frac{\pi^2  + {a^2}}{3}s^2
-\frac{(\pi^2 + {a^2})(2\pi^2 + 3{a^2})}{45} s^4
+\frac{(\pi^2 + {a^2})(\pi^2 + 2{a^2})(3\pi^2 + 5{a^2})}{945} s^6
-\frac{(\pi^2 + {a^2})^2(\pi^2 + 4{a^2}) }{4050} s^7
+\cdots,\!\!\!
\eea
\Rrule
\narrowtext
\noindent
which is in accord with the expansion (\ref{Es}), (\ref{detexp}).

In Fig.1 we show plots of the level spacing distributions 
\linebreak[2] $P(s)$ 
for various $a$ such that $\e^{-\frac{\pi^2}{a}}\ll 1$, 
obtained by numerically solving (33) 
under the boundary condition (\ref{bc}).
\begin{figure}
\epsfxsize=258pt
\begin{center}
\leavevmode
$\!\!\!\!\!\!\!\!\!\!$
\epsfbox{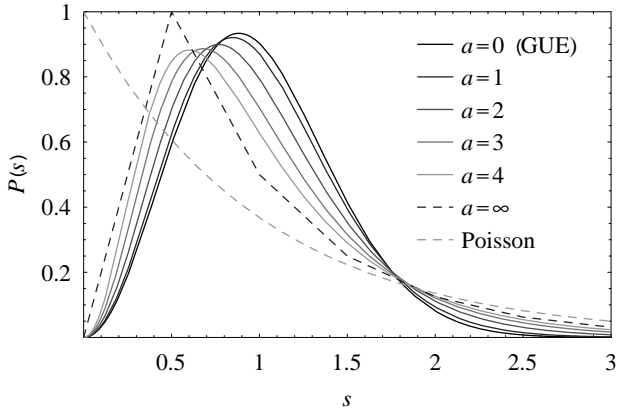}
\end{center}
\caption{
Level spacing distributions $P(s)$ of the kernel (\ref{KMNS}).
The limiting distribution for $a\rightarrow \infty$
(ref.[30], eq.(85)) and the Poisson distribution
are plotted for comparison.}
\end{figure}
\noindent
These distributions are indeed hybrids of 
the rescaled Wigner-Dyson distribution 
$P(s) \sim s^\beta$ ($\beta=2$) for $s {< \atop \sim} 1/a$
and the Poisson distribution 
$P(s) \sim \e^{-{\rm cst.}s}$ for $s {> \atop \sim} 1/a$.
The extension of our result to the case of orthogonal ensembles
($\beta=1$),
which correspond to the Anderson model, will be reported
elsewhere. 

I thank H. Widom for helpful correspondences, 
which is indispensable to this work.
I also thank
A. Kamenev, E. Kanzieper, V. Kravtsov and K. Muttalib
for discussions and comments.
The work of SMN is supported in part by 
JSPS Research Fellowships for Young Scientists,
by the Grant-in-aid No.411044
from the Ministry of Education, Science and Culture, Japan,
by Nishina Memorial Foundation,
and by NSF Grant No.PHY94-07194.

\widetext

\begin{references}
\bibitem[\dagger]{fn1}
The subscript reg means its regular part, i.e. with
$\delta$-functional peaks at coincident $x_i$'s subtracted.
\bibitem[\ddagger]{com}
This is not trivial because ref.\cite{TW4} has used
$\p=p^*$ and $\q=q^*$ which follow from the
analytic properties of its component functions,
$\phi(-x)=\phi(x)^*$ and $\psi(-x)=\psi(x)^*$.
\end{references}
\end{document}